\documentclass[aps,prl,showpacs,twocolumn,superscriptaddress]{revtex4}
\usepackage{graphics}
\usepackage{graphicx}
\usepackage{amsmath}
\usepackage{epsfig}
\usepackage{comment} 

\begin{document}

\title{Clamping of ferroelectric and ferromagnetic domain walls in conical spiral magnets}

\author{Andrea~Scaramucci}
\affiliation{Zernike Institute for Advanced Materials, University of Groningen, The Netherlands}
\author{Thomas~A.~Kaplan}
\affiliation{Department of Physics and Astronomy and Institute for Quantum Sciences, Michigan State University,
East Lansing, Michigan 48824, USA}
\author{Maxim~Mostovoy}
\affiliation{Zernike Institute for Advanced Materials, University of Groningen, The Netherlands}

\date{\today}

\begin{abstract}
We study the structure of domain walls in multiferroic magnets with the conical spiral ordering. We formulate a
simple spin model which has a conical spiral ground state in absence of magnetic anisotropies. We find a
transition from the regime where ferromagnetic and ferroelectric domain walls are clamped to the regime where
they are decoupled and derive a continuum model describing rotation of the spiral plane at the domain wall. The
importance of these results for the switching phenomena observed in CoCr$_2$O$_4$ is discussed.
\end{abstract}



\maketitle

{\it Introduction:} The simultaneous breaking of time-reversal and inversion symmetries, which gives rise to coupling between electric dipoles and magnetic moments in multiferroic materials, is observed in a number of
frustrated magnets with unusual spin orders and involves  interesting fundamental physics \cite{CheongNatMat2007}.
The practical motivation for studying these materials is the search for ways to control magnetic patterns by an
applied voltage and to manipulate electric polarization by magnetic field.

The magnetically-induced flip of the spontaneous electric polarization was first observed in nickel boracide
\cite{AscherJAP1966}, where the coexistence of a ferroelectric state with an antiferromagnetic ordering
exhibiting a linear magnetoelectric effect, gives rise to the magnetization, $\mathbf{M}$, linearly coupled to
the spontaneous electric polarization, $\mathbf{P}$ \cite{SannikovJETP1977}. More recently, the
90$^{\circ}$-rotation of electric polarization in an applied magnetic field was discovered  in rare earth
manganites with a spiral magnetic ordering \cite{KimuraNature2003}. The cycloidal spin spiral breaks inversion symmetry and induces electric polarization parallel to the spiral plane and normal to the spiral wave vector 
\cite{KatsuraPRL2005,SergienkoPRB2006,MostovoyPRL2006}. An applied magnetic field results in the
90$^{\circ}$-flop of the spiral plane with a concomitant rotation of $\mathbf{P}$. The giant peak in dielectric
constant of DyMnO$_3$ observed at the polarization flop transition \cite{GotoPRL2004} was recently related to
the motion of the 90$^{\circ}$ multiferroic domain walls in ac electric field \cite{KagawaPRL2009}. The
magnetically-induced polarization flip was also observed in HoMn$_2$O$_5$, which in an applied magnetic field
toggles between two different multiferroic phases with opposite orientations of $\mathbf{P}$ \cite{HurNature2004}.

Similar polarization reversals were recently found in the CoCr$_2$O$_4$ spinel, which shows a conical spiral
ordering and is both ferroelectric and ferromagnetic \cite{YamasakiPRL2006,ChoiPRL2009}. What is surprising about
these polarization flips is the absence of linear coupling between $\mathbf{M}$ and $\mathbf{P}$ in
CoCr$_2$O$_4$. The magnetization and polarization in this material are induced by two different magnetic
transitions: the magnetization appears below 95K as a result of the collinear ferrimagnetic ordering of Co and
Cr spins, while the electric polarization is induced by the spiral spin ordering that sets in at 27K. The
interaction between the uniform and spiral components of the conical spiral favors the spiral plane (and hence
the induced $\mathbf{P}$) orthogonal to $\mathbf{M}$,  but it cannot constrain the direction of rotation of
spins in the spiral (sometimes called chirality or handedness), which determines the sign of
$\mathbf{P}$. For a given $\mathbf{M}$, the spiral states with opposite handedness and opposite
polarizations are degenerate and both can be stabilized by electric field cooling \cite{YamasakiPRL2006,ChoiPRL2009}.

In Ref.~\cite{YamasakiPRL2006} it was suggested that the coupling between the magnetization and polarization in
CoCr$_2$O$_4$ occurs at domain walls, where $\mathbf{M}$ and $\mathbf{P}$ change sign simultaneously. The
coexisting ferrimagnetic and spiral orders imply the existence of at least two types of domain walls: the
ferromagnetic domain wall separating domains with opposite magnetization and the chiral or handed domain wall
where the direction of spin rotation and the induced electric polarization changes sign. The polarization
reversals in CoCr$_2$O$_4$ can be explained by clamping of ferromagnetic and chiral (ferroelectric) domain
walls.

In this Letter we study the structure of domain walls in conical spiral magnets and the clamping of
ferromagnetic (FM) and ferroelectric (FE) domain walls. The straightforward numerical simulation of the domain walls in
spinels is impossible, as the domain wall width, $w$, is large compared to the period of the spiral $l$. Since
maintaining the relations, $a \ll l \ll w \ll L$, where $a$ is the distance between neighboring spins and $L$ is
the system size, is crucial for studying the clamping of the FM and FE domain walls, we
devised a simple model of the conical spiral state, which is amenable to numerical simulations.  To our
knowledge, this is the first simple and rigorously soluble model where a conical spiral is stabilized without
magnetic anisotropy. The first model where this physics was found, but only variationally, is the cubic spinel
with nearest-neighbor AB and BB interactions, a complex structure with 6 coupled conical
spirals \cite{KaplanPhilMag2007}, which incidentally, has provided the basis for understanding the observed
behavior of several chromites  \cite{menyuk,KaplanPhilMag2007,YamasakiPRL2006,ChoiPRL2009}.

\begin{figure}[h]
    \centering
\includegraphics[width=0.48\textwidth]{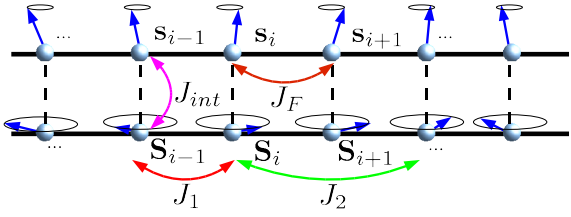}
    \caption{The system of coupled ferromagnetic and frustrated chains that shows a conical spiral ground state.}
    \label{fig:Model}
\end{figure}

{\it The model:} The model consists of two coupled chains of classical spins [see Fig.~\ref{fig:Model}]: one
with the FM coupling $J_{F}<0$ between neighboring  spins and another with the competing FM nearest-neighbor interaction $J_{1}<0$ and the antiferromagnetic next-nearest-neighbor interaction
$J_{2}>0$, which for $\frac{J_2}{\vert J_1 \vert} > \frac{1}{4}$ favors a simple (flat) spiral ordering with the wave vector $Q=\arccos \frac{\vert J_{1} \vert}{4 J_{2}}$. The Hamiltonian of the model is
\begin{eqnarray}
H =  \sum_{n=1}^L  \!\!&\!\!\![\!\!\!&\!\!
J_{F} \mathbf{s}_n \cdot \mathbf{s}_{n+1 } + J_{1} \mathbf{S}_{n} \cdot \mathbf{S}_{n+1}
 + J_{2}  \mathbf{S}_{n} \cdot \mathbf{S}_{n+2}
\nonumber \\
&\!\!+\!\!& J_{\rm int} \mathbf{s}_{n}\cdot\mathbf{S}_{n}],
\label{eqn:model}
\end{eqnarray}
where $\mathbf{s}_i$ and $\mathbf{S}_i$ are spins in, respectively, the ``ferromagnetic" and ``spiral"
chain, the spins being unit vectors.

When a weak interchain coupling $J_{\rm int}$ between the two chains is turned on, the ``ferromagnetic" chain
will acquire a small spiral component, while spins in the ``spiral" chain will rotate in the plane orthogonal to
the magnetization vector in the ``ferromagnetic" chain with a small canting in the magnetization direction. In
fact, such a conical spiral state,
\begin{equation}
\left\{
\begin{array}{rcl}
\mathbf{s}_{n} &=& s_{\perp} \left[\cos (Q n + \phi) \mathbf{e}_{1} +   \sin (Q n + \phi) \mathbf{e}_{2} \right] + s_{\parallel} \mathbf{e}_{3}, \\  \\
\mathbf{S}_{n} &=& S_{\perp} \left[\cos (Q n + \phi) \mathbf{e}_{1} +   \sin (Q n + \phi) \mathbf{e}_{2} \right]+ S_{\parallel} \mathbf{e}_{3},
\end{array}
\right.
\label{eqn:sol}
\end{equation}
where $\left(\mathbf{e}_{1},\mathbf{e}_{2},\mathbf{e}_{3} \right)$ is an orthogonal basis and $s_{\parallel}^2 +
s_{\perp}^2 = S_{\parallel}^2 + S_{\perp}^2 = 1$, is the ground state of Eq.(\ref{eqn:model}) in a wide range of
the model parameters.

This can be rigorously proven using the generalized Luttinger-Tisza method  and the idea of ``forced degeneracy'' \cite{LyonsPR1962,KaplanPhilMag2007}. In this method one minimizes the exchange energy, which is a
quadratic function of spins [see Eq.(\ref{eqn:model})], replacing the constraint on every site,
$\mathbf{S}_{n}^2=\mathbf{s}_{n}^2 = 1$ (the ``strong constraints"),  by the ``generalized weak constraint'',
$\sum_{n=1}^{L} \left(\mathbf{S}_{n}^2 + \alpha^2 \mathbf{s}_{n}^2\right) = L (1 + \alpha^2)$, where $\alpha$ is a real constant. If the minimal-energy spin configuration for some $\alpha$ also satisfies the strong constraints, it is a ground state of
Eq.(\ref{eqn:model}).

In momentum space the energy (\ref{eqn:model}) can be written,
\begin{equation}
 H = \sum_{q}
\mathbf{V}_{q}^{\dagger}
\left( \begin{array}{cc} J_1 \cos q + J_2 \cos 2q & \frac{J_{\rm int}}{2\alpha} \\ \frac{J_{\rm int}}{2\alpha} & \frac{J_F}{\alpha^2}\cos q  \end{array} \right) \mathbf{V}_{q},
\label{eqn:iM}
\end{equation}
where $\mathbf{V}_{q} = \left( \begin{array}{c} \mathbf{S}_{q} \\ \alpha \mathbf{s}_{q} \end{array} \right)$.
In some region of coupling constants it is possible to find $\alpha$ such that the lowest eigenvalue $\lambda(q,\alpha)$  has two degenerate global minima at $q=0$ and $q=Q$. In this case a solution for the weak
constraint is a linear combination of the two degenerate eigenvectors. The conical spiral state Eq.(\ref{eqn:sol})
belongs to the class of such solutions and satisfies the strong constraints.

\begin{figure}[htbp]
\centering
\includegraphics[width=0.45\textwidth]{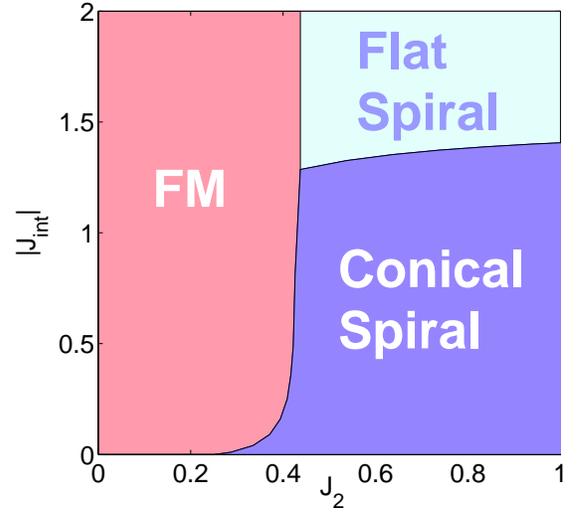}
\caption{The phase diagram of the two-chain model Eq.(\ref{eqn:model}) for $J_F= -0.75$ and $J_{\rm int} < 0$.  All exchange constants are measured in units of $\vert J_1 \vert = 1$.}
\label{fig:PD}
\end{figure}

The phase diagram of the model Eq.(\ref{eqn:model}) in the $J_2$-$J_{\rm int}$ plane is shown in Fig.~\ref{fig:PD}. 
For sufficiently strong interchain interaction $\mathbf{s}_{n} \| \mathbf{S}_{n}$ and the system becomes equivalent to a single chain showing either the FM or the flat spiral ordering. The vertical line separating these two states is $4 J_2 = \vert J_1\vert + \vert J_F \vert$, while the border line between the flat and conical spiral states is given by 
$
\frac{J_{\rm int}}{J_F} = 2 + \frac{\vert J_F \vert - \vert J_1 \vert}{2 J_2}.
$
This analytically obtained phase diagram was confirmed by numerical simulations of the two-chain model. 

The rotation of spins in the conical and flat spirals induces an electric polarization $\mathbf{P} \propto \mathbf{Q} \times \mathbf{e}_3$, where $\mathbf{Q}$ is the spiral wave vector \cite{KatsuraPRL2005,SergienkoPRB2006,MostovoyPRL2006}. In the conical spiral state the polarization vector is orthogonal to the magnetization $\mathbf{M} \| \mathbf{e}_3$ \cite{YamasakiPRL2006}.

{\it Domain walls:} Next we discuss the structure of domain walls in the conical spiral state. In the isotropic
spin model Eq.(\ref{eqn:model}) the width of the FM domain wall is of the order of the system size,
$w \sim L$, in which case the normal to the spiral plane follows the slowly rotating uniform magnetization, so
that $\mathbf{P}$ rotates together with $\mathbf{M}$. 
To make the width of the domain wall finite, we add to the exchange energy the magnetic anisotropy term on the
sites of the ferromagnetic chain,
\begin{equation}
H_{\rm a} = - \Delta \sum_n(s^{z}_{n})^2,
\label{eq:ani}
\end{equation}
which for $\Delta>0$ favors magnetization along the $z$ axis. The width of the FM domain is then $w \sim \pi \sqrt{\frac{\vert J_F \vert}{2\Delta}}$.

To ensure the presence of a FM domain wall in the system, we used the anti-periodic boundary
condition for spin projection on the easy axis in the ferromagnetic chain, $s_{n+L}^z =- s_n^z$. For all other
spin components we used open boundary conditions to prevent the enforced commensurability of the conical spiral state by finite size effects. Despite the simplicity of the model, its numerical simulations are computationally demanding, as frustrated spin interactions result in a rather complex structure of the domain wall. We find the
lowest-energy spin configuration for systems with up to 250 sites in each chain using the parallel tempering
method \cite{Swendsen,Earl}, which makes possible sampling of a large region of the phase space with multiple
local minima. 

\begin{figure}[ht]
    \centering
        \includegraphics[width=0.48\textwidth]{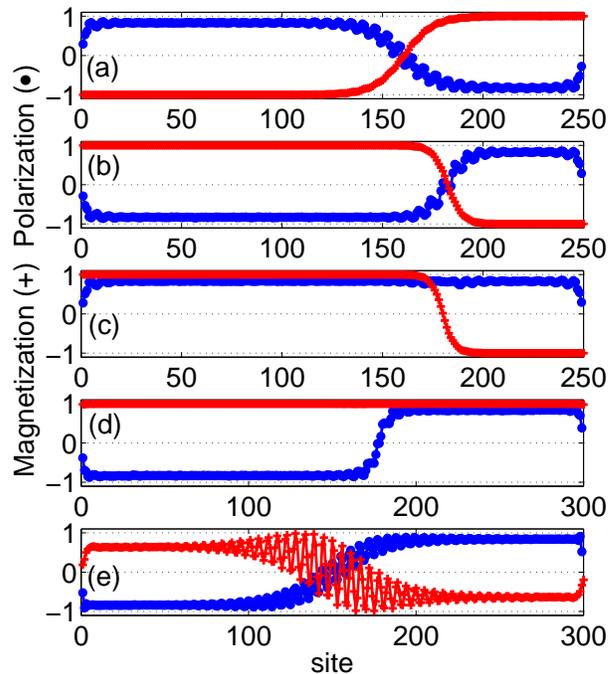}
    \caption{(Color online) The magnetization $s^{z}_{n}$ (red crosses) and the polarization $P_{n+1/2} = (\mathbf{S}_{n} \times \mathbf{S}_{n+1})^z$ (blue circles) in the domain wall. The first three panels show the minimal-energy state when a FM domain wall is enforced by boundary conditions for $J_2=0.5$, $J_F=-4$, $J_{\rm int}=-0.1$ and three different values of the magnetic ansitropy: (a) $\Delta=0.01$, (b) $\Delta=0.04$ (FM+FE domain wall) and (c)  $\Delta=0.05$ (FM domain wall). The other two panels show the ground state when a FE domain wall is enforced by an electric field $+E(-E)$ applied at the left(right) end of the spiral chain for: (d) $J_2=0.5$, $J_F=-4$,    $J_{\rm int}=-0.1$, and $\Delta=0.01$ (FE domain wall) and (e) $J_2=1$, $J_F=-1$, $J_{\rm int}=-1.5$, and $\Delta=0.01$ (FE + FM domain wall).  All exchange constants are measured in units of $\vert J_1 \vert = 1$. }
\label{fig:DWD}
\end{figure}

By varying the anisotropy strength $\Delta$ and, hence, the domain wall width, we observe a transition from the
regime where FM and FE domain walls are clamped to the regime where they are decoupled.
Figures~\ref{fig:DWD}(a), (b) and (c) show the magnetization along the easy axis direction in the ``ferromagnetic'' chain,
$s^z_{n}$, and the $y$-projection of the local polarization,
\begin{equation}
P_{n+1/2}\propto(\mathbf{S}_n \times\mathbf{S}_{n+1})^z,
\label{eq:polarization}
\end{equation}
induced in the ``spiral'' chain for two values of
the anisotropy constant $\Delta$ (we assume that the chains are parallel to the $x$ axis).

For relatively wide FM domain walls  the direction of the rotation of spins in the spiral changes
sign across the domain wall [see Fig.~\ref{fig:DWD}(a) and (b)], corresponding to the clamping of FM and FE domain walls responsible for the switching phenomena in CoCr$_2$O$_4$. When the thickness of the FM domain wall becomes smaller, the spins keep rotating in the same direction across the wall, so that the induced $\mathbf{P}$ has the same sign on both sides and the wall is purely ferromagnetic [see Fig.~\ref{fig:DWD}(c)].

Whether the FM domain wall will induce a FE domain wall or not depends on the balance between the interchain exchange energy, $E_{\rm int}$, and the energy cost of the FE domain wall, $E_{\rm FE}$. Expressions for the energies simplify in the limit $\vert J_F \vert \gg \vert J_{\rm int} \vert$, when $s_\perp \ll 1$ and the FM domain wall plays a role of the rotating magnetic field of magnitude $\vert J_{\rm int} \vert$ applied to the spiral chain. If, furthermore, the spiral wave vector, given by $\sin^2 \frac{Q}{2} = \frac{4 J_2 - \vert J_1\vert}{8 J_2}$, is small, the energy of the FE domain wall, where the spiral plane rotates at the rate $\sim \frac{\pi}{w} \ll Q$  together with the applied field, is
\begin{equation}
E_{\rm FE} \sim J_2 Q^2\left[1 - 2S_{\parallel}^2 \right] \left(\frac{\pi}{w}\right)^2 w,
\end{equation}
where $S_{\parallel} = \frac{\vert J_{\rm int} \vert}{J_2 Q^4}$, while the interchain energy is
\begin{equation}
E_{\rm int} \sim - \frac{1}{2} S_{\parallel} \vert J_{\rm int} \vert w.
\end{equation}
For $S_{\parallel} > \frac{1}{\sqrt{2}}$, the energy of the spiral decreases when its plane rotates and the FE and FM domain walls are clamped. This surprising result is explained by the fact that the spiral energy in the momentum space has a local {\em maximum} at $q = 0$, so that the rotation of the uniform component, $S_{\parallel} \mathbf{e}_{3}$, results in an energy decrease, which for $S_{\parallel} > \frac{1}{\sqrt{2}}$ exceeds the reduction due to the rotation of the spiral plane.  

\begin{figure}[h]
    \centering        \includegraphics[width=0.48\textwidth]{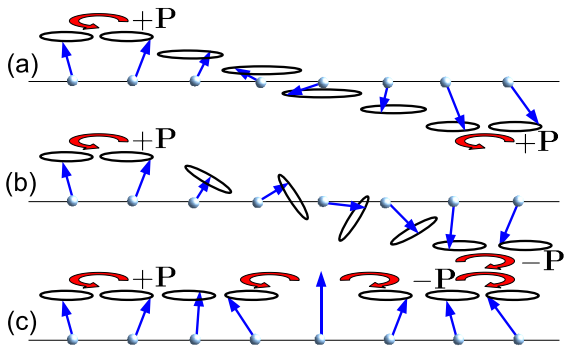}
    \caption{The sketch of the purely FM domain wall in the conical spiral magnet, where $\mathbf{M}$ changes sign while $\mathbf{P}$ does not (a), the FM+FE domain wall that reverses both $\mathbf{M}$ and $\mathbf{P}$ (b), and the FE domain wall, where the direction of spin rotation in the spiral is reversed, while $\mathbf{M}$ is unchanged (c).}
    \label{fig:artistic}
\end{figure}

For a nearly flat spiral, $S_{\parallel} \ll 1$, the dimensionless parameter determining the domain wall structure is $\lambda = \frac{Q w S_{\parallel}}{\pi \sqrt{2}} = 
\frac{2 \vert J_{\rm int}\vert}{\vert J_{1} \vert Q^3} \sqrt{\frac{\vert J_F\vert}{\Delta}}$: for 
$\lambda \gtrsim 1$, $E_{\rm FE} + E_{\rm int} < 0$ and the FM and FE walls are clamped, while for $\lambda \lesssim 1$ the domain wall is purely ferromagnetic. The shape of the domain wall for the nearly flat spiral can be found from the continuum limit of Eq.(\ref{eqn:model}):
\begin{equation}
E  =
\frac{\vert J_{\rm int} \vert}{\sqrt{2}Q}
\int\!\!dv
\left
[\left(\frac{d\Theta}{dv}\right)^2 + \sin^{2}(\Theta-\theta)\right] + const,
\label{eq:continuum}
\end{equation}
where the angles $\theta$ and $\Theta$ describe the rotation, respectively, of the magnetization and the spiral plane around an axis orthogonal to ${\hat z}$, and $v$ is the dimensionless coordinate along the chain, such
that the domain wall thickness is $\sim \lambda$. The second term in Eq.(\ref{eq:continuum}) is the interaction energy between the ferromagnetic and spiral subsystems. Minimizing $E$, one can show that for $\lambda \gg 1$ the spin rotation axis follows closely the direction of the magnetization: $\Theta - \theta \approx \lambda^{-2} \theta^{\prime\prime}\left(\frac{v}{\lambda}\right)$ [see Fig.~\ref{fig:artistic}b]. For $\lambda \ll 1$ the orientation of the spiral plane is only slightly perturbed in the vicinity of the FM domain wall: $\max(\vert \Theta \vert) \propto \lambda^2$, while $S_{\parallel} \approx \frac{\vert J_{\rm int} \vert}{J_2 Q^4} \cos \theta$ changes sign across the domain wall [see Fig.~\ref{fig:artistic}a].

Figures~\ref{fig:DWD}(d) and (e) show the minimal-energy spin configurations for the case when an electric field $+E(-E)$, applied at the left(right) end of the ``spiral" chain, stabilizes a FE domain wall, while the boundary conditions for spins are open. Since spins in the ``spiral chain" have no anisotropy of their own, the width of the FE domain wall is determined by the anisotropy in the FM chain and the interaction between the chains. Depending on the strength of this effective anisotropy, we find two different types of FE domain wall: the one where the spiral plane flips, as in the FE domain walls shown in Fig.~\ref{fig:DWD} and Fig.~\ref{fig:artistic}(b), and the one where the direction of spin rotation reverses, while the orientation of the spiral plane remains unchanged [see Fig.~\ref{fig:artistic}(c)].

The clamping of FM and FE domain walls is non-reciprocal:
the FE domain wall shown in Fig.~\ref{fig:DWD}(d) does not reverse $\mathbf{M}$, even though for the  
same set of parameters the FM domain wall induces the FE one [see Fig.~\ref{fig:DWD}(a)]. This non-reciprocity, originating from a difference in energies of the FM and FE domain walls seems to be a general phenomenon and may explain why the magnetization reversal in multiferroic GdFeO$_3$ results in an almost complete reversal of  $\mathbf{P}$, while the electrically-induced changes in $\mathbf{M}$ are very small \cite{TokunagaNatMat2009}.    

Spirals in frustrated magnets have a period of 10-20\AA~($l \sim 13.4$\AA~for CoCr$_2$O$_4$), while the typical width of a FM domain wall is an order of magnitude larger. Furthermore, in cubic CoCr$_2$O$_4$ the lowest-order magnetic anisotropy is of fourth order, which makes the domain wall width even larger, resulting in the perfect clamping of the FE and FM domain walls \cite{quartic}.

In conclusion, we presented (a) the first simple and rigorously soluble isotropic model with conical
ordering and (b) numerical and analytical studies of the structure of domain walls in a simple model
of multiferroic magnets with conical spiral ordering. We found a transition from the regime where ferromagnetic
and ferroelectric domain walls are clamped, to the regime where they are decoupled and identified the
dimensionless parameter that controls the transition. The clamping of ferroelectric and ferromagnetic domain
walls explains the apparent conservation of $\mathbf{P} \times \mathbf{M}$ (the ``toroidal
moment''~\cite{YamasakiPRL2006})  and the magnetically-induced polarization reversals observed in the CoCr$_2$O$_4$. 

\begin{acknowledgments}
This work was supported by the Thrust II program of the Zernike Insitute for Advanced Materials and by the Stichting voor Fundamenteel Onderzoek der Materie (FOM). TAK acknowledges helpful discussions with S. D. Mahanti.
\end{acknowledgments}

\end{document}